\documentclass[12pt]{article}
\usepackage{epsfig}
\usepackage{amssymb}
\usepackage{amsmath}
\usepackage{amsfonts}
\usepackage{graphicx}
\usepackage{mathrsfs}
\usepackage[dvips]{color}
\usepackage{multirow}

\newcommand{\R}{\mathbb{R}}
\newcommand{\C}{\mathbb{C}}

\newcommand{\fa}{\mathfrak{a}}
\newcommand{\fb}{\mathfrak{b}}
\newcommand{\fc}{\mathfrak{c}}

\newcommand{\fn}{{\mathfrak{n}}}

\newcommand{\fz}{\mathfrak{z}}

\newcommand{\bv}{\mathbf{v}}

\newcommand{\bA}{\mathbf{A}}
\newcommand{\bB}{\mathbf{B}}

\newcommand{\bM}{\mathbf{M}}
\newcommand{\bN}{\mathbf{N}}

\newcommand{\bT}{\mathbf{T}}

\newcommand{\cB}{\mathcal{B}}

\newcommand{\cM}{\mathcal{M}}

\newcommand{\cP}{\mathcal{P}}

\newcommand{\cS}{\mathcal{S}}
\newcommand{\cT}{\mathcal{T}}

\newcommand{\be}{\begin{equation}}
\newcommand{\ee}{\end{equation}}
\newcommand{\bea}{\begin{eqnarray}}
\newcommand{\eea}{\end{eqnarray}}

\newcommand{\ed}{\end{document}}

\newcommand{\bi}{\begin{itemize}}
\newcommand{\ei}{\end{itemize}}

\newcommand{\bce}{\begin{center}}
\newcommand{\ece}{\end{center}}

\newcommand{\sE}{\mathscr{E}}

\newcommand{\sH}{\mathscr{H}}

\newcommand{\sM}{\mathscr{M}}

\newcommand{\RE}{{\rm Re}}
\newcommand{\IM}{{\rm Im}}

\begin{document}

\title{Point Interactions, Metamaterials, and $\cP\cT$-Symmetry}

\author{Ali~Mostafazadeh\thanks{E-mail address:
amostafazadeh@ku.edu.tr}\\[6pt]
Departments of Mathematics and Physics, Ko\c{c} University,\\ 34450 Sar{\i}yer,
Istanbul, Turkey}

\date{ }
\maketitle

\begin{abstract}

We express the boundary conditions for TE and TM waves at the interfaces of an infinite planar slab of homogeneous metamaterial as certain point interactions and use them to compute the transfer matrix of the system. This allows us to demonstrate the omnidirectional reflectionlessness of Veselago's slab for waves of arbitrary wavelength, reveal the translational and reflection symmetry of this slab, explore the laser threshold condition and coherent perfect absorption for active negative-index metamaterials, introduce a point interaction modeling phase-conjugation, determine the corresponding antilinear transfer matrix, and offer a simple proof of the equivalence of Veselago's slab with a pair of parallel phase-conjugating plates. We also study the connection between certain optical setups involving metamaterials and a class of $\cP\cT$-symmetric quantum systems defined on wedge-shape contours in the complex plane. This provides a physical interpretation for the latter.
\vspace{2mm}

\vspace{2mm}

\end{abstract}

\section{Introduction}

Material with negative permittivity and permeability have unusual electromagnetic properties \cite{meta-review}. This observation was originally made by Veselago in 1968 \cite{veselago-1968}, but it was not until the year 2000 that it began to receive the attention of other physicists and become a focus of intensive research \cite{pendry-2000,meta-invisible}. The interest in this type of metamaterials was triggered and reenforced by two major developments. First, they were no long hypothetical substances but could be realized in laboratories \cite{meta-exp}. Second, they were shown to have remarkable applications in producing perfect lenses \cite{pendry-2000} and invisibility cloaks \cite{meta-invisible}.

\begin{figure}
	\begin{center}
	\includegraphics[scale=.8]{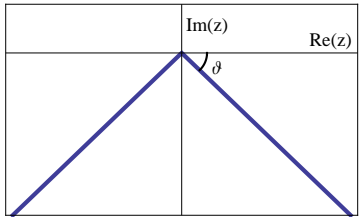}
	\caption{(Color online) A wedge-shaped contour in complex plane.}
	\label{fig1}
	\end{center}
	\end{figure}

The purpose of the present article is to explore some of the basic properties of metamaterials using the machinery of point interactions and transfer matrices. Specifically, we consider the scattering problem for transverse electric (TE) and transverse magnetic (TM) waves interacting with an infinite planar slab of homogeneous metamaterial. We identify the effect of the electromagnetic boundary conditions with the presence of certain point interactions and use them to obtain an explicit expression for the transfer matrix of the slab. This in turn allows us to investigate the invisibility properties and symmetries of the Veselago slab and derive explicit expressions for the laser threshold \cite{silfvast,pra-2011a} and antilasing conditions \cite{antilasing,CPA-Longhi} for active metamaterial slabs. Similarly we address the problem of modeling the phenomenon of phase conjugation \cite{phase-conjugatation} using certain antilinear point interactions and provide a very simple description of the equivalence of a Veselago slab to a pair of parallel phase-conjugating plates \cite{maslovski}. Next, we examine a recently studied spectral problem \cite{behrndt} that arises as a simplification of a metamaterial cloaking ring model \cite{bouchitte}. We offer a physical interpretation of this problem in terms of the transverse modes of a rectangular waveguide half-filled with a metamaterial with permittivity $-1$ and permeability $+1$, and establish its relation to the spectral problem for a $\cP\cT$-symmetric infinite square well potential defined on a wedge-shaped contour $\Lambda$ of the type shown Fig.~\ref{fig1}. We can express it in the form
    \be
    v(\fz):=\left\{\begin{array}{cc}
    0&{\rm for}~|\RE(\fz)|\leq \ell,\\
    \infty&{\rm otherwise},
    \end{array}\right.
    \ee
where $\fz$ marks a point on $\Lambda$ and $\ell$ is a positive real parameter. The quantum system associated with this potential was studied a decade ago in the context of obtaining a real description of $\cP\cT$-symmetric systems defined on complex contours \cite{jpa-2005a}. This was a crucial step towards the application of the formalism of pseudo-Hermitian quantum mechanics \cite{jpa-2004-review} for such systems. It was also the key to the subsequent results on the relation between $\cP\cT$-symmetric wrong-sign quartic potential, $v(x)=-x^4$, and the usual quartic anharmonic potential, $v(x)=x^4+\alpha x^2$, \cite{jones}. Here we relate $\cP\cT$-symmetric infinite square well potentials defined on wedge-shaped contours to the physics of metamaterials.

\section{Metamaterial Slabs and Point Interactions}

Consider an infinite planar slab $\hat\cS$ of thickness $L$ that is filled with a homogeneous (meta)material and placed in vacuum. If we take a cartesian coordinate system where the boundaries of $\hat\cS$ corresponds to the $z=0$ and $z=L$ planes, we can write the permittivity $\varepsilon$ and permeability $\mu$ of this system according to
    \begin{align}
    &\varepsilon=\left\{\begin{array}{ccc}
    \hat\varepsilon\,\varepsilon_0& {\rm for}& z\in[0,L],\\
    \varepsilon_0& {\rm for}& z\notin[0,L],
    \end{array}\right.
    &&\mu=\left\{\begin{array}{ccc}
    \hat\mu\,\mu_0& {\rm for}& z\in[0,L],\\
    \mu_0& {\rm for}& z\notin[0,L],
    \end{array}\right.
    \end{align}
where $\varepsilon_0$ and $\hat\varepsilon$ are the permittivity of the vacuum and the relative permittivity of the slab, and $\mu_0$ and $\hat\mu$ are the permeability of the vacuum and the relative permittivity of the slab, respectively.

\begin{figure}
	\begin{center}
	\includegraphics[scale=.35]{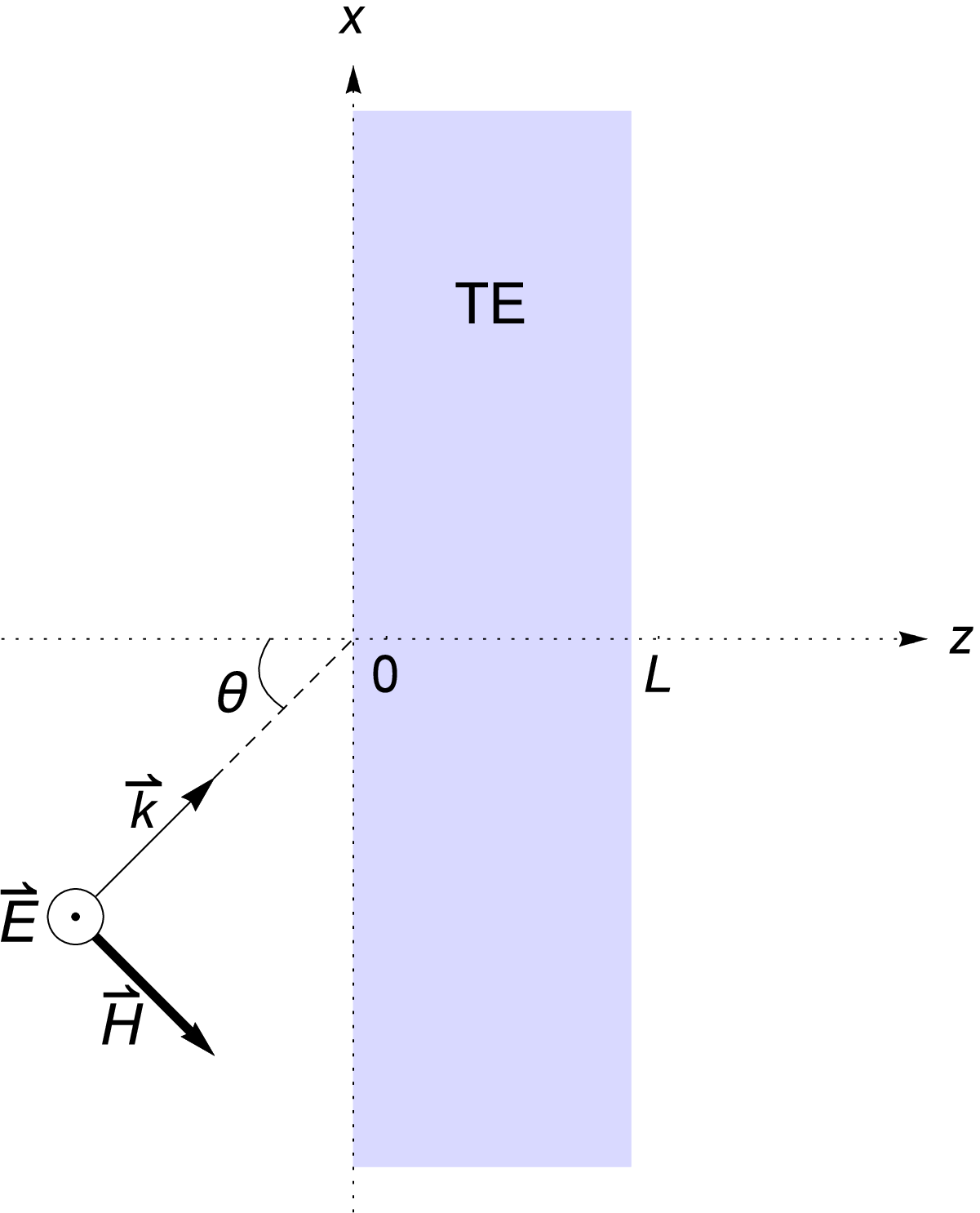}~~~~~~~
    \includegraphics[scale=.35]{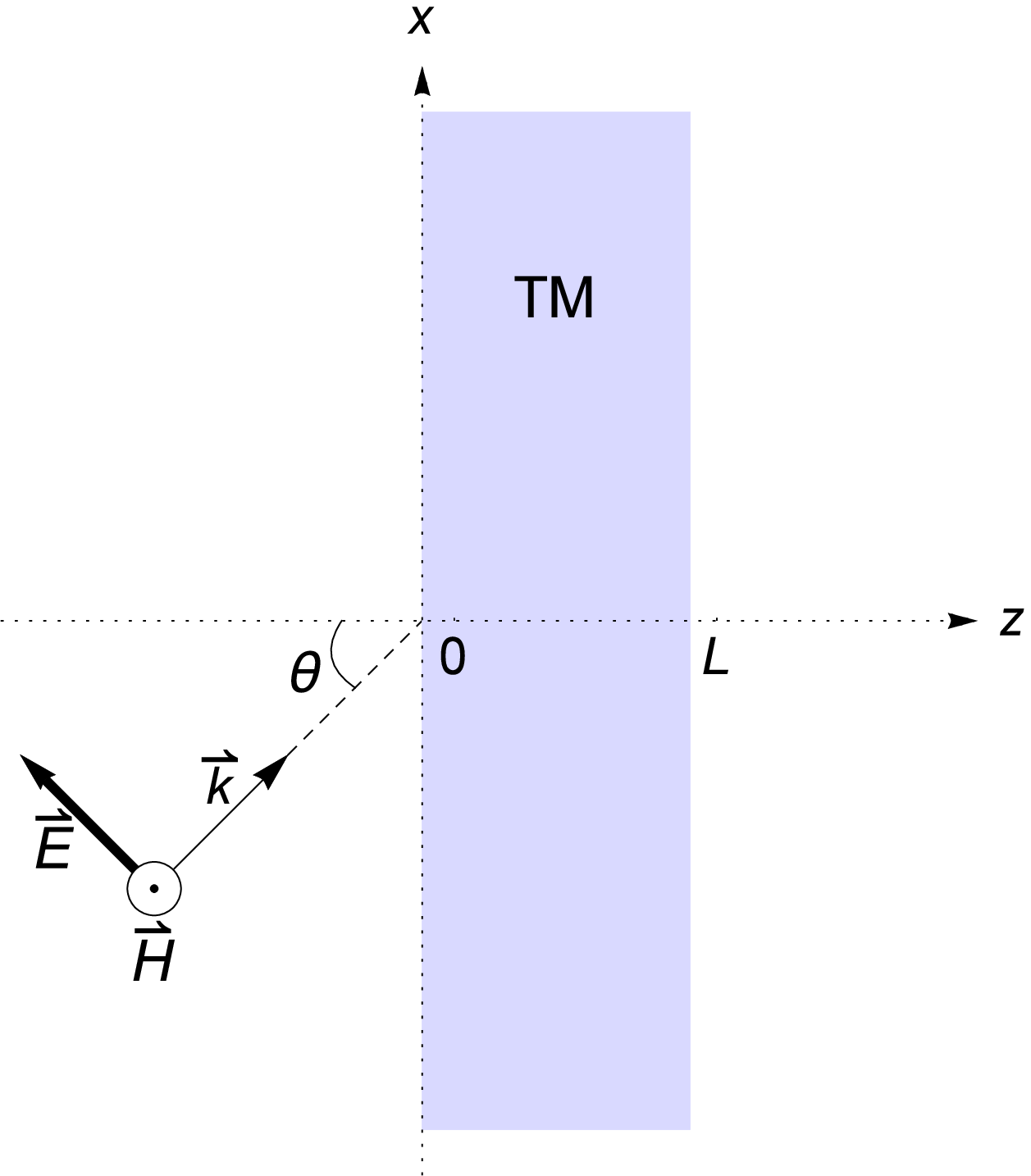}
	\caption{(Color online) Graphs of the TE (on the left) and TM (on the right) waves incident on a slab of (meta)material (the colored region).}
	\label{fig2}
	\end{center}
	\end{figure}

Now, consider a time-harmonic TE (respectively TM) wave incident on $\hat\cS$ such that the electric field $\vec{E}$ of the TE (magnetic field $\vec{H}$ of the TM) wave lies along the $y$-axis and the  wavevector $\vec{k}$ makes an angle $\theta$ with the $z$-axis as shown in Fig.~\ref{fig2}. Then,
    \begin{align}
    &\vec{k}=k_x\,\vec e_x+k_z\,\vec e_z,
    && k_x:=k\sin\theta, &&k_z:=k\cos\theta,
    \label{wavevector}
    \end{align}
where $k:=|\vec{k}|=\omega/c$ is the wavenumber, $\omega$ is the angular frequency of the wave,  $c:=1/\sqrt{\varepsilon_0\,\mu_0}$ is the speed of light in vacuum, and $\vec{e}_x, \vec{e}_y,$ and $\vec{e}_z$ are the unit vectors along the positive $x$-, $y$-, and $z$-axes, respectively.

With these conventions in place, we can write the TE and TM solutions of the Maxwell equations,
    \begin{align}
    &\vec\nabla\cdot(\varepsilon\,\vec E)=0, && \vec\nabla\cdot(\mu\,\vec H)=0,
    \label{max1}\\
    &\varepsilon\,\partial_t \vec E=\vec{\nabla}\times\vec H, &&
    \mu\,\partial_t \vec H=-\vec{\nabla}\times\vec E,
    \end{align}
as
    \begin{align}
    &{\rm TE:}~\left\{\begin{aligned}
    &\vec E=e^{i(k_x x-\omega t)}\,\sE(z)  \vec{e}_y, \\
    &\vec H:=(c\,\mu)^{-1}e^{i(k_x x-\omega t)}
    \left[ik^{-1}\sE'(z)\vec{e}_x+\sin\theta\,\sE(z)\vec{e}_z\right],\end{aligned}\right.
    \label{TE=}\\[6pt]
    &{\rm TM:}~\left\{\begin{aligned}
    &\vec E:=-(c\,\varepsilon)^{-1}e^{i(k_x x-\omega t)}
    \left[ik^{-1}\sH'(z)\vec{e}_x+\sin\theta\,\sH(z)\vec{e}_z\right],\\
    &\vec H=e^{i(k_x x-\omega t)}\sH(z)\vec{e}_y.
    \end{aligned}\right.
    \label{TM=}
    \end{align}
Here $\sE$ and $\sH$ are solutions of the Helmholtz equation,
    \be
    \psi''(z)+k^2(\fn^2-\sin^2\theta)\psi(z)=0,~~~~~~z\notin\{0,L\},~~~~~~\fn^2:=\varepsilon\,\mu\,c^2,
    \label{Helm-eqn}
    \ee
which satisfy the matching conditions,
    \begin{align}
    &\sE(z_0^+)=\sE(z_0^-), && \sE{\,'}(z_0^+)=\alpha^{\rm TE}_{z_0}\sE{\,'}(z_0^-),
    \label{match-E}\\
    &\sH(z_0^+)=\sH(z_0^-), && \sH{\,'}(z_0^+)=\alpha^{\rm TM}_{z_0}\sH{\,'}(z_0^-),
    \label{match-H}
    \end{align}
at $z_0=0,L$, for every function $f:\R\to\R$, $f(z_0^{-/+})$ stands for the limit of $f(z)$ as $z$ tends to $z_0$ from the left/right, and
    \begin{align*}
    &\alpha^{\rm TE}_{z_0}:=\frac{\mu(z_0^+)}{\mu(z_0^-)},
    &\alpha^{\rm TM}_{z_0}:=\frac{\varepsilon(z_0^+)}{\varepsilon(z_0^-)}.
    \end{align*}
For the system we consider,
    \begin{align}
    &\fn^2=\left\{
    \begin{array}{ccc}
    \hat\varepsilon\,\hat\mu & {\rm for} & z\in[0,L],\\
    1 & {\rm for} & z\notin[0,L],\end{array}\right.
    && \alpha^{\rm TE}_{0}=\frac{1}{\alpha^{\rm TE}_{L}}=\hat\mu,
    && \alpha^{\rm TM}_{0}=\frac{1}{\alpha^{\rm TM}_{L}}=\hat\varepsilon.
    \label{match3}
    \end{align}

The term metamaterial refers to the unusual situations where the real part of $\hat\varepsilon$, $\hat\mu$, or both have a negative sign. The best known example is a Veselago metamaterial, where
$\hat\varepsilon=\hat\mu=-1$, \cite{veselago-1968}. As seen from the above equations, all the interesting properties of the latter follows from the corresponding matching conditions (\ref{match-E}) and (\ref{match-H}) at the boundaries. This observation has motivated the authors of \cite{maslovski} to argue that the same effects can be realized using a pair of parallel phase-conjugating plates placed in vacuum. In the remainder of this section we consider a more general situation where $\hat\varepsilon$ and $\hat\mu$ are arbitrary complex numbers with a negative real part and try to construct an equivalent system consisting of a slab $\check\cS$ of ordinary material and a pair of plates that realize the relevant boundary conditions in such a way that the permittivity $\check\varepsilon$ and permeability $\check\mu$ of $\check\cS$ differ from those of $\hat\cS$ only by the sign of the real part of $\check\varepsilon$ and $\check\mu$, i.e.,
    \begin{align}
    &\begin{aligned}
    &\RE(\check\varepsilon)=-\RE(\hat\varepsilon),\\
    &\IM(\check\varepsilon)=\IM(\hat\varepsilon),
    \end{aligned}
    &&\begin{aligned}
    &\RE(\check\mu)=-\RE(\hat\mu),\\
    &\IM(\check\mu)=\IM(\hat\mu).
    \end{aligned}
    \label{check}
    \end{align}

Let us view the Helmholtz equation (\ref{Helm-eqn}) as the time-independent Schr\"odinger equation
    \be
    -\psi''(z)+v(z)\psi(z)=k_z^2\psi(z),
    \label{sch-eq}
    \ee
for the potential
    \be
    v(z):= k^2(1-\fn^2)=k_z^2(1-\tilde\fn^2),
    \label{v=}
    \ee
where
    \be
    \tilde\fn:=\pm\sec\theta\sqrt{\fn^2-\sin^2\theta},
    \label{tilde-n}
    \ee
and the undetermined sign in this relation is fixed by the requirement that the real part of $\tilde\fn$ and $\fn$ have identical sign, i.e.,
    \be
    \RE(\tilde\fn)\RE(\fn)\geq 0.
    \label{tilde-n-sign}
    \ee
Then the presence of the above-noted plates corresponds to the addition of certain point interactions to the potential (\ref{v=}). Notice also that (\ref{v=}) is, in general, a barrier potential that vanishes outside $[0,L]$ and has a (possibly) complex hight $k^2(1-\hat\varepsilon\hat\mu)$ .

The determination of the point interactions that map the boundary conditions for the metamatrial slab $\hat\cS$ to those of $\check\cS$ requires the knowledge of the effect of the addition of a point interaction to a finite-range scattering potential. This is most conveniently acquired using the transfer matrix formulation of scattering theory.

\section{Transfer Matrix for a Metamaterial Slab}

For a scattering potential $v(z)$ which decays sufficiently rapidly as $z\to\pm\infty$, every solution of the Schr\"odinger equation~(\ref{sch-eq}) has the following asymptotic form.
    \[\psi(x)\to A_\pm e^{ik_z z}+B_\pm e^{-ik_z z}~~{\rm for}~~z\to\pm\infty.\]
The transfer matrix of $v(z)$ is by definition the $2\times 2$ matrix $\bM$ satisfying
    \[\left[\begin{array}{c}
    A_+\\ B_+\end{array}\right]=\bM\,\left[\begin{array}{c}
    A_-\\ B_-\end{array}\right].\]
Let us recall that the left- and right-incident scattering solutions of (\ref{sch-eq}) are defined by the asymptotic boundary conditions:
    \begin{align}
    &\psi_l(z)\to N_l\times \left\{\begin{array}{ccc}
    e^{ik_zz}+R^l e^{-ik_zz} & {\rm for} & z\to -\infty,\\
    T^l e^{ik_zz} & {\rm for} & z\to\infty,\end{array}\right.\\
    &\psi_r(z)\to N_r\times \left\{\begin{array}{ccc}
    T^r e^{-ik_zz} & {\rm for} & z\to -\infty,\\
    e^{-ik_zz}+R^r e^{ik_zz} & {\rm for} & z\to\infty,\end{array}\right.
    \end{align}
where $N_{l/r}$ are normalization constants, and $R^{l/r}$ and $T^{l/r}$ are respectively the left/right reflection and transmission amplitudes. These are related to the entries and determinant of $\bM$ according to
    \begin{align}
    &R^l=-\frac{M_{21}}{M_{22}}, &&
    R^r=\frac{M_{12}}{M_{22}}, &&
    T^l=\frac{\det\bM}{M_{22}}, &&
    T^r=\frac{1}{M_{22}}.
    \label{R-T=}
    \end{align}
Using the Wronskian identities satisfied by the solutions of (\ref{sch-eq}), one can show that $\det\bM=1$, \cite{jpa-2009,sanchez}. This relation holds for both real and complex scattering potentials. It implies that the left and right transmission amplitudes are equal. Because of this we use $T$ for $T^{l/r}$, whenever we consider a scattering problem obeying $\det\bM=1$.

An important property of transfer matrices is their composition rule. This states that if $v(z)$ is the sum of two potentials $v_-(z)$ and $v_+(z)$ such that the support of $v_-$ lies to the left of that of $v_+$, i.e., there is some $a\in\R$, $v_-(z)=0$ for $z>a$ and $v_+(z)=0$ for $z<a$, then the transfer matrix of $v(z)$ has the form $\bM=\bM_+\bM_-$, where $\bM_\pm$ is the transfer matrix of $v_\pm(z)$.

The above description of the transfer matrices, scattering solutions, and reflection and transmission amplitudes applies also for the scattering due to a point interaction. The latter is a zero-range interaction whose effect is to impose matching conditions for $\psi(z)$ and $\psi'(z)$ at a discrete set of points $z_0$ in $\R$. For example consider a point interaction with a single center $z_0$ that is defined by
    \be
    \left[\begin{array}{c}
    \psi(z_0^+)\\ \psi'(z_0^+) \end{array}\right]=\bB_{z_0}\,\left[\begin{array}{c}
    \psi(z_0^-)\\ \psi'(z_0^-)\end{array}\right],
    \label{pt-1}
    \ee
where $\bB_{z_0}$ is an arbitrary invertible matching matrix \footnote{There are also other types of point interaction. See for example those considered in \cite{cgk}.}. Then for $z\neq z_0$ the system corresponds to a free particle and we can easily determine its transfer matrix. This gives \cite{jpa-2011}
    \be
    \bM_{z_0}=\bN_{z_0}^{-1}\bB_{z_0}\,\bN_{z_0},
    \label{M-pt1}
    \ee
where
    \be
    \bN_{z_0}:=\left[\begin{array}{cc}
    e^{ik_zz_0} & e^{-ik_zz_0}\\
    i k_z e^{ik_zz_0} & -i k_z e^{-ik_zz_0}\end{array}\right].
    \label{M-pt}
    \ee
An immediate consequence of (\ref{M-pt1}) is the existence of anomalous point interactions that violate the condition $\det\bM=1$. Clearly they correspond to matrices $\bB_{z_0}$ that do not have a unit determinant \cite{jpa-2011}.

We can identify the electromagnetic boundary conditions (\ref{match-E}) and (\ref{match-H}) with a pair of point interactions with center $z_0=0,L$ and diagonal matching matrices:
    \begin{align}
    &\bB^{\rm TE/TM}_{z_0}:=\left[\begin{array}{cc}
    1 & 0 \\ 0 & \alpha_{z_0}^{\rm TE/TM}\end{array}\right].
    \label{beta=}
    \end{align}
Substituting this relation in (\ref{M-pt1}), we can express the corresponding transfer matrix as
    \be
    \bM^{\rm TE/TM}_{z_0}=\bM_{z_0}(\alpha^{\rm TE/TM}),
    \label{TEM-M}
    \ee
where
    \begin{align}
    &\bM _{z_0}(\alpha):=\frac{1}{2}\left[\begin{array}{cc}
    1+\alpha  & (1-\alpha )e^{-2ik_zz_0} \\
    (1-\alpha )e^{2ik_zz_0} & 1+\alpha \end{array}\right].
    \label{M=10}
    \end{align}
Clearly, $\det\bM^{\rm TE/TM}_{z_0}=\det\bB^{\rm TE/TM}_{z_0}=\alpha_{z_0}^{\rm TE/TM}$. Therefore, electromagnetic boundary conditions generally correspond to anomalous point interactions. However, often these come in pairs in such a way that their contribution to the determinant of the transfer matrix of the system as a whole cancel and we again find a total transfer matrix that has a unit determinant.

Next, we confine our attention to the TE waves. Let $\bM^{\rm TE}$ denote the transfer matrix for the TE waves and $\bM_\star$ be its value for a fictitious slab $\cS_\star$ of the same size and position that is filled with material with permittivity $\varepsilon_\star:=\varepsilon_0\hat\varepsilon\hat\mu$ and permeability $\mu_\star:=\mu_0$. It is easy to see that $\bM_\star$ coincides with the transfer matrix of a rectangular barrier potential (\ref{v=}) which admits a simple explicit expression \cite{ap-2014}. It is convenient to write it in the form
    \be
    \bM_\star=\boldsymbol{\cM}(\tilde\fn\,k_z L,\tilde\fn_+,\tilde\fn_-),
    \label{M-star=}
    \ee
where $\tilde\fn_\pm:=(\tilde\fn\pm\tilde\fn^{-1})/2$, and for all $\fa,\fb,\fc\in\C$,
    \be
    \boldsymbol{\cM}(\fa,\fb,\fc):=\left[
    \begin{array}{cc}
    \left(\cos\fa+i\,\fb\sin\fa\right)e^{-ik_zL} &
    i\,\fc \sin\fa\,e^{-ik_zL}\\[6pt]
    -i\,\fc\sin\fa\,e^{ik_zL} &
    \left(\cos\fa-i\,\fb\sin\fa\right)e^{ik_zL}
    \end{array}\right].
    \label{boldcal-M}
    \ee
Furthermore, in view of the composition property of transfer matrices, and Eqs.~(\ref{match3}), (\ref{TEM-M}), (\ref{M-star=}), and (\ref{boldcal-M}), we have the following remarkably simple result.
    \be
    \bM^{\rm TE}=\bM_{L}^{\rm TE}\,\bM_\star\,\bM^{\rm TE}_{0}=
    \boldsymbol{\cM}(\tilde\fn\,k_z L,\fn^{\rm TE}_+,\fn^{\rm TE}_-),
    \label{MTE=}
    \ee
where $\fn^{\rm TE}_\pm:=(\tilde\fn^2\pm\hat\mu^2)/2\hat\mu\,\tilde\fn$. Using a similar analysis we can compute the transfer matrix $\bM^{\rm TM}$ for the TM waves. This yields
    \be
    \bM^{\rm TM}=\boldsymbol{\cM}(\tilde\fn\,k_z L,\fn^{\rm TM}_+,\fn^{\rm TM}_-),
    \label{MTM=}
    \ee
where $\fn^{\rm TM}_\pm:=(\tilde\fn^2\pm\hat\varepsilon^2)/2\hat\varepsilon\,\tilde\fn$. Equations~(\ref{boldcal-M}) -- (\ref{MTM=}) imply $\det\bM^{\rm TE}=\det\bM^{\rm TM}=1$. Hence we have reciprocal transmission; $T^l=T^r=:T$.

For a Veselago slab, which we denote by $\cS_v$, we have $\hat\varepsilon=\hat\mu=-1$, $\fn^2=\tilde\fn^2=1$, and (\ref{MTE=}) and (\ref{MTM=}) give
    \be
    \bM^{\rm TE}=\bM^{\rm TM}=\bM_v:=\left[\begin{array}{cc}
    e^{-2 i k_z L} & 0\\
    0 & e^{2 i k_zL}
    \end{array}\right].
    \label{M-veselago}
    \ee
Therefore, in view of (\ref{R-T=}), $\cS_v$ is omnidirectionally reflectionless for waves of arbitrary wavenumber. But it is not generally invisible, because it induces a phase shift in the transmitted wave; $T=e^{-2ik_zL}$.

Let us also note that according to (\ref{boldcal-M}), (\ref{MTE=}), and (\ref{MTM=}), $\hat\cS$ is reflectionless for TE/TM modes whenever $\fn_-^{\rm TE/TM}=0$. This corresponds to
the Brewster angles $\theta_b$ give by
    \be
    \cos\theta_b=\left\{\begin{array}{cc}
    |\hat\mu|\sqrt{\frac{\fn^2-1}{\hat\mu^2-1}} &\mbox{for TE waves},\\[9pt]
    |\hat\varepsilon|\sqrt{\frac{\fn^2-1}{\hat\varepsilon^2 -1}} &\mbox{for TM waves},
    \end{array}\right.
    \ee
which has the same form for $\check S$, i.e., the Brewster angles for metamaterials coincide with those of the usual material.

Another remarkable property of a Veselago slab is that its transfer matrix and consequently its scattering features are invariant under space translations and reflections about any plane parallel to the slab. This follows from the fact that under a space translation, $z\stackrel{T_a}{\longrightarrow} z-a$, and the space reflection (parity), $z\stackrel{\cP}{\longrightarrow}-z$, transfer matrices transform as
    \begin{align}
    &\bM\stackrel{T_a}{\longrightarrow}\bT_a^{-1}\bM\,\bT_a,
    && \bM\stackrel{\cP}{\longrightarrow}\boldsymbol{\sigma_1}\bM^{-1}\boldsymbol{\sigma_1},
    \end{align}
where
    \begin{align}
    &\bT_a:=e^{iak_z\boldsymbol{\sigma_3}}=\left[\begin{array}{cc}
    e^{iak_z}& 0\\0 & e^{-iak_z}\end{array}\right],
    && \boldsymbol{\sigma_1}=\left[\begin{array}{cc}
    0& 1\\1 & 0\end{array}\right],
    && \boldsymbol{\sigma_3}=\left[\begin{array}{cc}
    1& 0\\0 & -1\end{array}\right].
    \label{bT=}
    \end{align}

Next, consider placing an arbitrary infinite planar slab $\cS$ between a pair of parallel Veselago slabs of thickness $L$. The space-translation invariance of Veselago slabs implies that the transfer matrix of both of the Veselago slabs have the form (\ref{M-veselago}). Therefore, if we use $\bM_{\cS}$ to denote the transfer matrix of $\cS$, then according to the composition property of the transfer matrices, the transfer matrix of the whole system of three slabs has the form
    \be
    \bM=\bM_v\bM_{\cS}\bM_v=\left[\begin{array}{cc}
    e^{-4i k_z L}(T_{\cS}-R_{\cS}^lR_{\cS}^r/T_{\cS}) & R_{\cS}^r/T_{\cS}\\
    -R_{\cS}^l/T_{\cS} & e^{4i k_z L}/T_{\cS}\end{array}\right].
    \label{VSV}
    \ee
Here we have made use of (\ref{R-T=}), assumed that $\det\bM_{\cS}=1$ (so that the left and right transmission amplitudes of $\cS$ coincide), and denoted the reflection and transmission amplitudes of $\cS$ by $R^{l/r}_{\cS}$ and $T_{\cS}$. Equation~(\ref{VSV}) shows that
no matter what the distance between the slabs are, the transmission amplitude of $\cS$ undergoes a phase shift, according to $T\to e^{-4ik_zL}T$, while its reflection amplitudes do not change. This marks the application of the Veselago slabs as omnidirectional phase shifters.

\section{Laser Threshold and CPA Conditions}

The condition that an optical potential supports a spectral singularity \cite{prl-2009} corresponds to what is known as the laser threshold condition in optics \cite{pra-2011a}. Because the time-reversal of this condition yields coherent perfect absorption (CPA) or antilasing \cite{antilasing,CPA-Longhi}, the knowledge of spectral singularities allows for the determination of the laser threshold and CPA conditions in a variety of optical setups \cite{SS1,SS2}. Reference~\cite{pra-2015a} makes use of this approach to provide a detailed description of the behavior of the TE and TM waves interacting with a nonmagnetic optically active slab of ordinary material. Because spectral singularities (respectively their time-reversal) correspond to the real zeros of the $M_{22}$ (respectively $M_{11}$) entry of the transfer matrix in the complex $k$-space \cite{prl-2009,CPA-Longhi,SS2}, we can easily use (\ref{boldcal-M}), (\ref{MTE=}) and (\ref{MTM=}) to do the same for the metamaterial slab $\hat\cS$. As a result we find the following compact expression for the laser threshold and CPA conditions.
    \be
    e^{2ik_z L\,\tilde\fn}=\left(\frac{\fn_+^{\rm TE/TM}+1}{\fn_+^{\rm TE/TM}-1}\right)^{\!\gamma},
    \label{LT-CPA}
    \ee
where
    \[\gamma:=\left\{\begin{array}{ccc}
    1 & \mbox{for~lasing},\\
    -1& \mbox{for~CPA}.\end{array}\right.\]

In order to derive a formula for the threshold gain (loss) coefficient and the phase condition for lasing (CPA), we introduce
    \begin{align*}
    &\varepsilon_1:=\RE(\hat\varepsilon), &&\varepsilon_2:=\IM(\hat\varepsilon),
    &&\mu_1:=\RE(\hat\mu), &&\mu_2:=\IM(\hat\mu),
    \end{align*}
so that
    \[\hat\varepsilon=\varepsilon_1+i\varepsilon_2,~~~~~~~~~~~~~\hat\mu=\mu_1+i\mu_2,\]
and confine our attention to situations where the ratio of $|\varepsilon_2|$ and $|\mu_2|$ to $|\varepsilon_1\mu_1-1|$ are several orders of magnitude smaller than $1$, i.e.,
    \[ \frac{|\varepsilon_2|+|\mu_2|}{|\varepsilon_1\mu_1-1|}\ll 1.\]
In this case, except for the incidence angles close to the Brewster's angle, we can ignore the quadratic and higher order terms in $\varepsilon_2$ and $\mu_2$, \cite{pra-2015a}. For example, in view of the fact that $\fn^2= \hat\varepsilon\hat\mu $, we can write
    \be
    \fn \approx \xi\left[\sqrt{\varepsilon_1\mu_1}+\frac{i}{2}(\varepsilon_1\mu_2+\mu_1\varepsilon_2)\right]
    =\xi\sqrt{\varepsilon_1\mu_1}+\frac{i}{2}\,\big(|\varepsilon_1|\mu_2+|\mu_1|\varepsilon_2\big),
    \label{n2=approx}
    \ee
where
    \[\xi:= \left\{\begin{array}{ccc}
    -1 & {\rm for} &  \varepsilon_1>0~{\rm and}~\mu_1>0,\\
    1 & {\rm for} &  \varepsilon_1<0~{\rm and}~\mu_1<0,\end{array}\right.\]
and we have taken note of the fact that for a metamaterial with $\varepsilon_1<0$ and $\mu_1<0$, $\RE(\fn)<0$, \cite{pendry-2000}.

Observe that (\ref{n2=approx}) is consistent with the fact that for a gain metamaterial, $\epsilon_2<0$ and $\mu_2<0$. To see this, consider a normally incident TE wave, where $\theta=0$, $k_z=k$, and $\tilde\fn=\fn$. Then, as the wave travels a distance $d$ inside the slab, the intensity of the electric field, namely $|\sE(z)|^2$, changes by a factor of $|e^{i\fn k d}|^2$. The gain/loss coefficient is identified with the quantity $g$ that allows us to write this factor in the form $e^{gd}$. Therefore, by virtue of (\ref{n2=approx}), it has the form
    \be
    g:=-2k\IM(\fn)=
    -k(|\varepsilon_1|\mu_2+|\mu_1|\varepsilon_2)~~{\rm for}~~\varepsilon_1\mu_1>0.
    \label{g=}
    \ee
Clearly, if $\epsilon_2<0$ and $\mu_2<0$, $g>0$ and the (meta)material is a gain medium \cite{Hess-2012}.

Next, we substitute (\ref{n2=approx}) in (\ref{tilde-n}) and use (\ref{tilde-n-sign}) and (\ref{g=}) to obtain
    \be
    \tilde\fn \approx \sec\theta(\xi\,\nu_1+i\nu_2)
    \label{tilde-n-approx}
    \ee
where
    \begin{align}
   &\nu_1:=\sqrt{\varepsilon_1\mu_1-\sin^2\theta},~~~
    &&\nu_2:=\frac{\nu_3}{2} \big( |\varepsilon_1|\mu_2+|\mu_1|\varepsilon_2 \big)=
    -\frac{\nu_3 g}{2k},~~~
    && \nu_3:=\frac{1}{\sqrt{1-\frac{\sin^2\theta}{\varepsilon_1\mu_1}}}.
    \label{nus}
    \end{align}
Inserting (\ref{tilde-n-approx}) in the expression for $\fn_+^{\rm TE/TM}$ and employing the result together with (\ref{nus}) and (\ref{tilde-n-approx}) in (\ref{LT-CPA}), we find the following laser threshold (CPA) and phase conditions
    \begin{align}
    &g\approx\frac{2\gamma}{\nu_3 L}\ln\left(\frac{\eta+1}{\eta-1}\right),
    &&k\approx\frac{2\gamma(\pi m+\phi)}{\nu_1 L},
    \label{Condis}
    \end{align}
where we have only retained the leading order terms in powers of $\varepsilon_2$ and $\mu_2$ and introduced
    \bea
    \eta&:=&\left\{\begin{array}{cc}
    \sec\theta\,\nu_1/|\mu_1|&\mbox{for TE waves},\\
    \sec\theta\,\nu_1/|\varepsilon_1|&\mbox{for TM waves},\end{array}\right. \\
    \phi&:=&\left\{\begin{array}{cc}
    \frac{\eta}{\eta^2-1}\left[\frac{1}{2\pi m}
    \ln\left(\frac{\eta+1}{\eta-1}\right)+\frac{\mu_2}{|\mu_1|}\right]&\mbox{for TE waves},\\[12pt]
    \frac{\eta}{\eta^2-1}\left[\frac{1}{2\pi m}
    \ln\left(\frac{\eta+1}{\eta-1}\right)+\frac{\varepsilon_2}{|\varepsilon_1|}\right]&\mbox{for TM waves},
    \end{array}\right.
    \label{phi=}
    \eea
and a mode number $m$ that takes positive integer values. Relations~(\ref{nus}) -- (\ref{phi=}) show that the sign of the real part of the permittivity and permeability does not enter the laser threshold and CPA conditions. This shows that as far as these conditions are concerned active metamaterials behave exactly like the usual active material.

\section{Metamaterials from Point Interactions and\\ Phase-Conjugation}

Consider the matching matrix (\ref{beta=}) that represents the boundary condition at the interfaces of the metamaterial slab $\hat\cS$ and vacuum. We can obtain the matching matrix $\check\bB^{\rm TE/TM}_{z_0}$ for the slab $\check\cS$ by replacing $\alpha_{z_0}^{\rm TE/TM}$ by $\check\alpha_{z_0}^{\rm TE/TM}$ in (\ref{beta=}) where
    \begin{align}
    &\check\alpha_{0}^{\rm TE}=1/\check\alpha_{L}^{\rm TE}=\check\mu=|\mu_1|+i\mu_2,
    &&\check\alpha_{0}^{\rm TM}=1/\check\alpha_{L}^{\rm TM}=\check\varepsilon=|\varepsilon_1|+
      i\varepsilon_2.
    \end{align}
It is not difficult to see that
    \be
    \bB^{\rm TE/TM}_{z_0}=\check\bB^{\rm TE/TM}_{z_0}\boldsymbol{\cB}^{\rm TE/TM}_{z_0},
    \label{B=BB}
    \ee
where
    \be
    \boldsymbol{\cB}^{\rm TE/TM}_{z_0}:=\left[\begin{array}{cc}
    1 & 0\\ 0 &-e^{2i\varphi_{z_0}}\end{array}\right],
    \label{meta-pt}
    \ee
and $\varphi_{z_0}$ is the principal argument (phase angle) of $\alpha_{z_0}^{\rm TE/TM}$, i.e. $e^{i\varphi_{z_0}}=\alpha_{z_0}^{\rm TE/TM}/|\alpha_{z_0}^{\rm TE/TM}|$. In other words,
    \[\varphi_{0}:=-\varphi_{L}=\left\{
    \begin{array}{cc}
    \arctan(\mu_2/\mu_1) & \mbox{for~TE waves},\\
    \arctan(\varepsilon_2/\varepsilon_1) & \mbox{for~TM waves}.\end{array}\right.\]

Equation~(\ref{B=BB}) suggests that if we view $\check\bB^{\rm TE/TM}_{z_0}$ and $\boldsymbol{\cB}^{\rm TE/TM}_{z_0}$ as defining certain point interactions, then we can identify the transfer matrix $\bM_{z_0}^{\rm TE/TM}$ for the point interaction given by $\bB^{\rm TE/TM}_{z_0}$ with the product of those associated with $\check\bB^{\rm TE/TM}_{z_0}$ and $\boldsymbol{\cB}^{\rm TE/TM}_{z_0}$. Denoting these with $\check\bM^{\rm TE/TM}_{z_0}$ and $\boldsymbol{\cM}^{\rm TE/TM}_{z_0}$, we have
    \[\bM_{z_0}^{\rm TE/TM}=\check\bM^{\rm TE/TM}_{z_0}\,\boldsymbol{\cM}^{\rm TE/TM}_{z_0}.\]
In light of (\ref{M-pt}), $\boldsymbol{\cM}^{\rm TE/TM}_{z_0}$ has the following explicit expression.
    \be
    \boldsymbol{\cM}^{\rm TE/TM}_{z_0}=
    e^{i\varphi}\left[\begin{array}{cc}
    -i\sin\varphi & e^{-2ik_zz_0}\cos\varphi\\
    e^{2ik_zz_0}\cos\varphi & -i\sin\varphi \end{array}\right].
    \label{cM=45}
    \ee
For a Veselago slab, $\check\varepsilon=\check\mu=1$, $\varphi=0$,
    \be
    \bM_{z_0}^{\rm TE/TM}=\boldsymbol{\cM}^{\rm TE/TM}_{z_0}=
    \left[\begin{array}{cc}
    0& e^{-2ik_zz_0}\\
    e^{2ik_zz_0}& 0 \end{array}\right],
    \label{cM=46}
    \ee
and the transfer matrix of the slab takes the form $\bM^{\rm TE/TM}=\bM_{L}^{\rm TE/TM}\bM_{0}^{\rm TE/TM}$. Substituting (\ref{cM=46}) in this equations we recover (\ref{M-veselago}).

Because the bulk transfer matrix $\bM_\star$ for our metamaterial slab $\hat\cS$ is the same as the one for $\check\cS$, this suggests that we can view $\hat\cS$ as $\check\cS$ supplemented with a pair of point interactions given by $\boldsymbol{\cB}^{\rm TE/TM}_{z_0}$ at its boundaries. In particular, a Veselago slab of thickness $L$ is equivalent to a pair of point interactions with center $z_0=0,L$ and matching conditions $\boldsymbol{\cB}^{\rm TE/TM}_{z_0}=\boldsymbol{\sigma}_3$. Therefore all the surprising properties of the metematerial slabs may be realized using certain surface interactions that take place at their boundaries. The same conclusion is reached by the authors of Ref.~\cite{maslovski} who argue that the properties of a Veselago slab are common to those of a pair of parallel phase-conjugating plates. These can also be modeled using an antilinear point interaction of the form
    \be
    \left[\begin{array}{c}
    \psi(z_0^+)\\
    \psi'(z_0^+)\end{array}\right]=
    \left[\begin{array}{c}
    \psi(z_0^-)^*\\
    -\psi'(z_0^-)^*\end{array}\right]=
    \boldsymbol{\sigma_3}\boldsymbol{\star}\left[\begin{array}{c}
    \psi(z_0^-)\\
    \psi'(z_0^-)\end{array}\right],
    \label{antilinear-pt}
    \ee
where $\boldsymbol{\star}$ stands for the antilinear operator of complex-conjugation of matrices; given a $2\times 2$ matrix $\bA$ and a column vector ($2\times 1$ matrix) $\bv$, we have $\boldsymbol{\star}\bA\bv=\bA^*\boldsymbol{\star}\bv=\bA^{\!*}\bv^{\!*}$.

We can purse the approach of \cite{jpa-2011} to determine the transfer matrix of (\ref{antilinear-pt}). This results in an antilinear matrix of the form
    \be
    \boldsymbol{\sM}_{z_0}:=\bN_{z_0}^{-1}\boldsymbol{\sigma_3}\bN_{z_0}^*\boldsymbol{\star}
    =\bT_{-z_0}\boldsymbol{\star},
    \label{anti-M}
    \ee
where $\bN_{z_0}$ and $\bT_a$ are respectively given by (\ref{M-pt1}) and (\ref{bT=}). It is easy to show that $\boldsymbol{\sM}_{z_0}^2$ coincides with the identity matrix. This is a manifestation of the fact that an overlapping pair of phase-conjugating plates should behave exactly like vacuum.

The composition property of transfer matrices implies that the transfer matrix of a pair of parallel phase-conjugating plates located at $z=0$ and $z=L$, is given by
    \be
    \boldsymbol{\sM}_{L}\boldsymbol{\sM}_{0}=\bT_{-L}\boldsymbol{\star}
    \bT_{0}\boldsymbol{\star}=\bT_{-L}=\left[\begin{array}{cc}
    e^{-2 i k_z L} & 0\\
    0 & e^{2 i k_zL}
    \end{array}\right],
    \ee
where we have made use of (\ref{anti-M}). The fact that the right-hand side of this relation coincides with the transfer matrix of Veselago's slab (\ref{M-veselago}) provides a straightforward proof of the equivalence of the latter with a pair of parallel phase-conjugating plates \cite{maslovski}.

\section{Metamaterials and $\cP\cT$-symmetry}

The point interactions defined by (\ref{meta-pt}) correspond to the matching conditions of the form
    \be
    \psi(z_0^-)=\psi(z_0^+),~~~~~~~~~~~~~~
    e^{-i2\vartheta}\psi'(z_0^-)=e^{i2\vartheta}\psi'(z_0^+),
    \label{PT-match}
    \ee
where
    \be
    \vartheta:=\frac{\pi}{4}-\frac{\varphi}{2}.
    \label{vartheta}
    \ee
It is quite remarkable that these matching conditions also appear in the study of a class of $\cP\cT$-symmetric spectral problems defined on the wedge-shaped complex contours of the form \cite{jpa-2005a}:
    \be
    \Gamma :=\big\{\fz-i\tan\vartheta\:|\fz|~\big|~\fz\in\R\big\}.
    \label{cantour}
    \ee
See also Fig~\ref{fig1}. It turns out that the solution of the time-independent Schr\"odinger equation,
    \[-\Psi''(\fz)+V(\fz)\Psi(\fz)=E\Psi(\fz),\]
defined in $L^2(\Gamma)$ is equivalent to the spectral problem defined in $L^2(\R)$ by the differential equation
    \be
    e^{2i\,\vartheta\,{\rm sgn}(z)}
    \left[-\psi''(z)+v(z)\psi(z)\right]=E\psi(z)~~~~~~~~z\in\R\setminus\{0\},
    \label{sch-eq-G}
    \ee
and the matching conditions~(\ref{PT-match}), where $v(z)$ is determined by $V(\fz)$ and $\vartheta$, \cite{jpa-2005a}.

\begin{figure}
	\begin{center}
	\includegraphics[scale=.6]{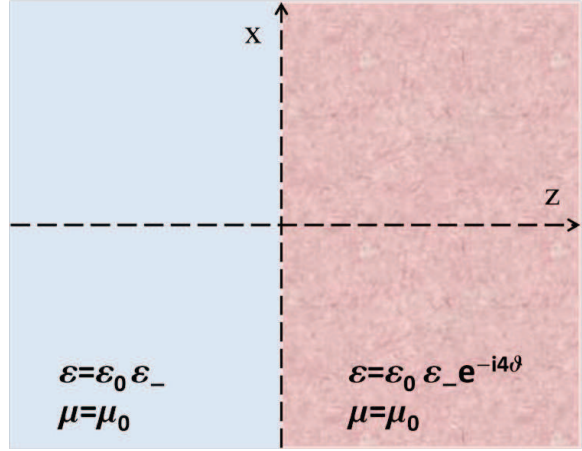}\vspace{-12pt}
	\caption{(Color online) An infinite volume of space. The blue region corresponding to $z<0$ consists of nonmagnetic ordinary material. The pink region corresponding to $z>0$ contains nonmagnetic (meta)material.}
	\label{fig3}
	\end{center}
	\end{figure}

Let us consider a volume of infinite extent that is separated into two regions by the $z=0$ plane and filled with nonmagnetic (meta)materials ($\mu=\mu_0$) such that the permittivity of the two regions differ only by a phase factor according to
    \begin{align}
    &\varepsilon=\left\{\begin{array}{ccc}
    \varepsilon_0\,\varepsilon_- &{\rm for} & z<0,\\
    e^{-4i\vartheta}\varepsilon_0\,\varepsilon_-&{\rm for} & z\geq 0,
    \end{array}\right.
    \end{align}
where $\varepsilon_-\in\R^+$ and $\vartheta\in[0,\pi/4]$. See Fig.~\ref{fig3}.

It is not difficult to see that the TM waves propagating along the $z$-axis in this volume are given by (\ref{TM=}) provided that we set $\theta=0$ and demand that $\sH(z)$ satisfies (\ref{PT-match}) and (\ref{sch-eq-G}) with $v(z)=0$ and $E:=k^2 e^{-2i\vartheta} \varepsilon_-$. In view of the results of \cite{jpa-2005a}, this identifies the behavior of these waves with that of the eigenfunctions of the $\cP\cT$-symmetric Hamiltonian operator for a free particle moving on the contour (\ref{cantour}). The special case $\varepsilon_-=1$ and $\vartheta=\pi/4$ corresponds to a volume half-filled with a nonmagnetic metamaterial with permittivity $-1$. In this case $\sH(z)$ solves the spectral problem:
	\begin{align}
	&-{\rm sgn}(z)\frac{d^2}{dz^2}\psi(z)=\lambda_z\psi(z),~~~~~z\neq 0,\\
	&\psi(0^+)=\psi(0^-),~~~~~\psi'(0^+)=-\psi'(0^-),
	\label{spec-1}
	\end{align}
with
	\be
	\lambda_z=-i E=- k^2 \varepsilon_-.
	\label{eg-va}
	\ee
    \begin{figure}[t]
	\begin{center}
    \includegraphics[scale=.6]{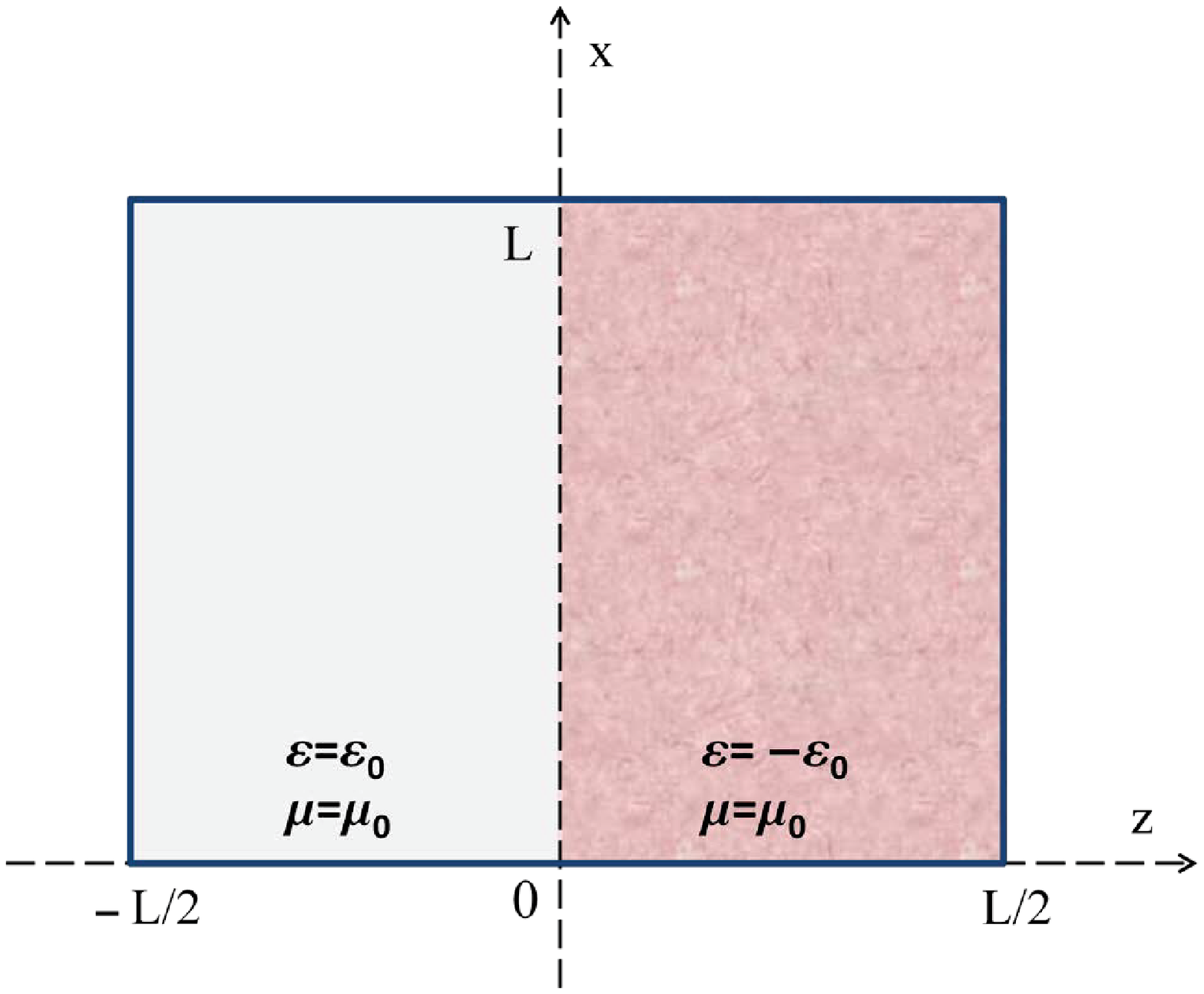}
	\caption{Cross section of a rectangular waveguide aligned along the $y$-axis and half-filled with a nonmagnetic metamaterial with permittivity $-\varepsilon_0$ (pink region).}
	\label{fig4}
	\end{center}
	\end{figure}
	
Next, consider confining this system to the region defined by $0\leq x\leq L$ and $-L/2\leq z\leq L/2$. This corresponds to waves propagating in a rectangular waveguide aligned along the $y$-axis. See Fig.~\ref{fig4}. If we  suppose that the guide has  perfectly reflecting walls and try to obtain its transverse modes, we should solve the spectral problem obtained by supplementing (\ref{spec-1}) with the Dirichlet boundary conditions at $z=\pm L/2$, i.e., set
	\be
	\psi(\pm L/2)=0.
	\ee
This determines the $z$-dependent part of the corresponding standing waves for the magnetic field $\vec H$. The $x$-dependent part is determined by the solutions of
	\be
	-\frac{d^2}{dx^2}\chi(x)=\lambda_x\chi(x),~~~~~\chi(0)=\chi(L)=0,
	\label{spec-02}
	\ee
in $[0,L]$. In other words, the standing magnetic waves for this system are given by the solutions of the spectral problem:	
	\begin{align}
	&-{\rm sgn}(z)\nabla^2 f(x,z)=\lambda f(x,z), \quad\quad
	x\in(0,L),~z\in(-\mbox{$\frac{L}{2}$},0)\cup(0,\mbox{$\frac{L}{2}$}),
	\label{spec3a}\\[6pt]
	&f(0,z)=f(L,z)=0,
	\quad\quad z\in[-\mbox{$\frac{L}{2}$},\mbox{$\frac{L}{2}$}],
	\label{spec3b}\\[6pt]
	&f(x,0^+)-f(x,0^-)=\partial_z f(x,0^+)+\partial_z f(x,0^-)=f(x,\pm\mbox{$\frac{L}{2}$})=0,
	\quad\quad x\in[0,L],
	\label{spec3c}
	\end{align}
which is defined on the square $\square:=\big\{\:(x,z)\:\big|\:0\leq x\leq L,\: |z|\leq L/2\;\big\}$.

The authors of Ref.~\cite{behrndt} consider the purely mathematical problem of whether (\ref{spec3a}) -- (\ref{spec3c})  can be cast into a spectral problem for a genuine self-adjoint operator acting in $L^2(\square)$. They answer this question in the affirmative and show that $0$ is an infinitely-degenerate eigenvalue of this operator and that the other eigenvalues are located symmetrically about $0$.  They also use the separability of (\ref{spec3a}) to determine the general form of the separable solutions, i.e., the standing waves of our waveguide in terms of products of the solutions of (\ref{spec-1}) and (\ref{spec-02}). The latter gives $\chi(x)=\sin(\pi n x/L)$ with $n=1,2,3,\cdots$, while the former can be expressed in terms of $\sin$ and $\sinh$ functions \cite{behrndt}.

As we noted above (\ref{spec-1}) is a special case of the $\cP\cT$-symmetric spectral problem defined by (\ref{sch-eq-G}), (\ref{PT-match}), and $\psi(\pm\frac{L}{2})=0$, which is treated in \cite{jpa-2005a}. As shown in this reference, for $\vartheta>\pi/4$, none of the corresponding eigenvalues is real. The $\cP\cT$-symmetry of the problem implies that they appear in complex-conjugate pairs. For $\vartheta=\pi/4$ which is relevant to our model, $E_0=0$ is the only real eigenvalue and again the rest of the eigenvalues form complex-conjugate pairs.
By analogy to its one-dimensional counterpart, we can also reveal the $\cP\cT$-symmetric nature of (\ref{spec3a}) -- (\ref{spec3c}). We can certainly write the differential equation give in (\ref{spec3a}) as
$L f=\tilde\lambda f$ where
	\be
	L:=-i\,{\rm sgn}(z)\nabla^2,~~~~~~~~~~~~~\tilde\lambda=i\lambda.
	\label{t-spec3a}
	\ee
Clearly $L$ is $\cP\cT$-symmetric, for $\cP f(x,z):=f(x,-z)$ and $\cT f(x,z):=f(x,z)^*$. Furthermore, $\cP\cT$ leaves the set of boundary conditions given in (\ref{spec3b}) and (\ref{spec3c}) invariant. Therefore, written in terms of $L$ and $\tilde\lambda$, (\ref{spec3a}) -- (\ref{spec3c}) is essentially a $\cP\cT$-symmetric spectral problem. The fact that the spectrum of (\ref{spec3a}) -- (\ref{spec3c}) is real and symmetric about zero is equivalent to the condition that the spectrum of  $L f=\tilde \lambda f$ defined via the same boundary conditions is purely imaginary and symmetric about the real axis. In other words the complex eigenvalues come in complex-conjugate pairs. This a characteristic feature of all $\cP\cT$-symmetric spectral problems.

This example also reminds us of the following rather obvious but mostly unappreciated consequence of the characterization theorems of Refs.~\cite{p12,p3}.
	\begin{itemize}
	\item[]{\bf Theorem:} {\em Let $A$ be a Hermitian (self-adjoint) operator acting in a separable Hilbert space and having a discrete spectrum. Then the anti-Hermitian operator $\tilde A:=iA$ is pseudo-Hermitian provided that the spectrum of $A$ is symmetric about $0$ and the eigenvalues of $A$ that are related by the reflection about $0$ have the same multiplicity.}
	\end{itemize}
By virtue of a result of \cite{p3}, we can further infer that the operator $\tilde A$ commutes with an antilinear involution, i.e., it possesses a generalized $\cP\cT$-symmetry \cite{jmp-2003}. Because the spectrum is not real, this symmetry is always broken.

\section{Concluding Remarks}

During the past fifteen years or so, the purely theoretical results of Veselago on unusual properties of metamaterials with negative permittivity and permeability have found experimental verifications. The efforts in this direction together with the seminal works of Pendry and his collaborators \cite{pendry-2000,meta-invisible} on the application of negative-index metamaterials in constructing perfect lenses and invisibility cloaks have made their study a major area of research in theoretical and experimental physics.

In this article we have employed the idea of point interactions together with the transfer matrix formulation of scattering theory to explore some of the basic properties of metamaterials. In particular, we have computed the transfer matrix of a general (meta)material slab both for TE and TM waves and used the result to demonstrate the omnidirectional reflectionlessness of the Veselago slab for waves of arbitrary wavenumber. We have shown that sandwiching an arbitrary slab between a pairs of Veselago slabs does not affect the scattering features of the original slab except that it shifts the phase of the transmitted waves. This marks the application of Veselago slabs as phase shifters.
We have also established the fact that the Brewster's angles and the laser threshold and CPA conditions do not depend on the sign of the real part of the permittivity and permeability of a (meta)material slab, if these have the same sign.

Another result we have reported in this article is a simple proof of the equivalence of a Veselago slab with a pair of parallel phase-conjugating plates \cite{maslovski}. This required the introduction of a class of antilinear point interactions. The transfer matrix of these interactions turn out to be antilinear matrices squaring to the identity matrix. These curious observations call for a comprehensive study of antilinear point interactions and their physical applications.

Finally, we have shown how certain configurations of metamaterials are related to a class of exotic $\cP\cT$-symmetric models defined on wedge-shaped complex contours. These were originally introduced for purely theoretical reasons \cite{jpa-2005a}. Our results provide a simple physical interpretation for these models.

\vspace{12pt}
\noindent{\em Acknowledgments:}  I am grateful to Kaan G\"uven and Sasan Hajizadeh for suggesting references on active metamaterials and helping me find and correct typos in the first draft of this article. This work has been supported by  the Scientific and Technological Research Council of Turkey (T\"UB\.{I}TAK) in the framework of the project no: 114F357, and by the Turkish Academy of Sciences (T\"UBA).

\ed